# A microprocessor based on a two-dimensional semiconductor

Stefan Wachter[1,*], Dmitry K. Polyushkin[1,*], Ole Bethge[2], and Thomas Mueller[1,‡]

[1] *Vienna University of Technology, Institute of Photonics, Gußhausstraße 27-29, 1040 Vienna, Austria*
[2] *Vienna University of Technology, Institute of Solid State Electronics, Floragasse 7, 1040 Vienna, Austria*

[*] These authors contributed equally to this work.
[‡] Corresponding author: thomas.mueller@tuwien.ac.at

**The advent of microcomputers in the 1970s has dramatically changed our society. Since then, microprocessors have been made almost exclusively from silicon, but the ever-increasing demand for higher integration density and speed, lower power consumption and better integrability with everyday goods has prompted the search for alternatives. Germanium and III-V compound semiconductors are being considered promising candidates for future high-performance processor generations [1] and chips based on thin-film plastic technology [2] or carbon nanotubes [3] could allow for embedding electronic intelligence into arbitrary objects for the Internet-of-Things. Here, we present a 1-bit implementation of a microprocessor using a two-dimensional semiconductor – molybdenum disulfide. The device can execute user-defined programs stored in an external memory, perform logical operations and communicate with its periphery. Importantly, our 1-bit design is readily scalable to multi-bit data. The device consists of 115 transistors and constitutes the most complex circuitry so far made from a two-dimensional material.**

Two-dimensional (2D) materials, such as semiconducting transition metal dichalcogenides (TMDs) [4, 5], black phosphorus [6], silicence [7] and others, are considered promising candidates for future generations of electronic circuits. Although it is currently not foreseen that 2D materials will replace silicon for mainstream digital electronics in the mid-term future, they offer a number of interesting properties that could lead to novel applications. Their ultrathin channel thickness provides improved electrostatic gate control and reduced short-channel effects [8, 9], which ultimately results in better geometric scaling behavior [10] and less power consumption. 2D semiconductors are also one of the leading candidates to enable tunnel field-effect transistors [11, 12], working with sub-threshold swing below 60 mV per decade and thus low supply voltage. Together, with their high mechanical flexibility and stability, optical transparency, and excellent optoelectronic properties [13], this could lead to energy efficient and flexible electronics [14–16].

The field of TMD-based electronics has progressed enormously during the past few years. Soon after the first realizations of bulk [17, 18] and monolayer [6] field-effect transistors (FETs), basic electronic circuits were demonstrated [19, 20]. Both n-type

(NMOS) [19, 20, 21] and complementary (CMOS) [22, 23] metal-oxide-semiconductor technologies have been developed and a good understanding of the FET device physics has been gained [24–26]. The work on devices has been paralleled by the development of growth techniques [27–30] for the large-scale fabrication of TMD films with good uniformity over the size of a wafer [30] and the development of technologies for transferring 2D materials onto bendable substrates [14–16]. Nevertheless, due to the plethora of challenges being faced in large-scale integration, previous work has so far been restricted to applications consisting of only a few transistors and with limited functionality. These challenges range from the necessity to match voltage levels and achieve high noise margins in cascaded logic stages to stringent requirements on device uniformity over millimeter size dimensions. In this work, we demonstrate the feasibility of using a 2D semiconductor to realize a complex digital circuit – a microprocessor.

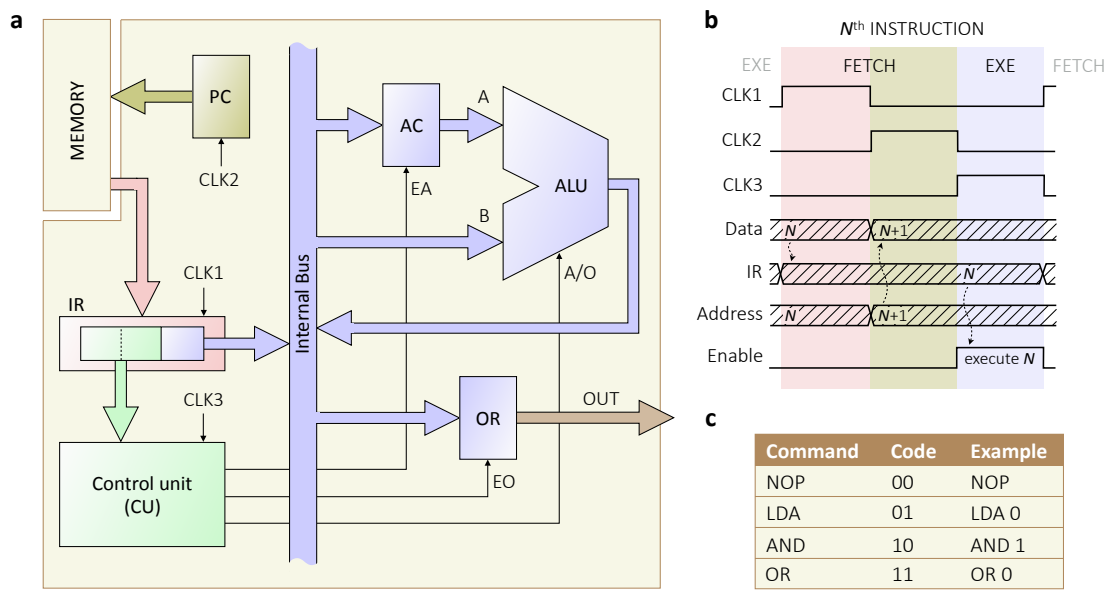

**Figure 1| Microprocessor architecture. (a)** Block diagram, showing the arithmetic logic unit (ALU) with inputs *A* and *B*, accumulator (AC), control unit (CU), instruction register (IR), output register (OR), and program counter (PC). Enable signals (*EA* and *EO*) and operation selection code (*A/O*) are supplied by the CU to the respective subunits. CLK signal generation and memory are implemented off-chip. **(b)** Timing diagram for the $N^{th}$ instruction cycle. During the *FETCH* sequence the content of the memory is loaded into the IR and the address, stored in the PC, is increased. During the *EXE* sequence the command, stored in the IR, is executed. **(c)** Instruction set of the microprocessor. *NOP* is the no-operation instruction; *LDA* transfers data from the memory into the AC; *AND* and *OR* perform logical operations.

Figure 1a depicts the architectural block diagram of our microprocessor. For demonstration purposes, we minimized transistor count and thus realized a device that operates on single-bit data only. We stress that this is not a fundamental limitation and the device is readily scalable to *N*-bit data, loosely speaking by connecting *N* of our devices in parallel. Although we reduced the architecture of our device to the essentials, it comprises all basic building blocks that are common to most microprocessors. In particular, these are: (i) An *arithmetic-logic unit* (ALU), that forms the heart of the processor and is, in general, capable of performing basic arithmetic and logical operations. For simplicity, we have implemented here only logical conjunction and disjunction operations. (ii) An *accumulator* (AC), which holds one of the operands to be supplied to the ALU. (iii) An *instruction register* (IR), that stores the content of the program memory currently being executed, where the most significant two bits contain the instruction itself and the third bit contains the data. (Although we retrieve the data directly from the program memory, our device can also process data stored in a separate data memory (Harvard architecture). In this case, the IR is supplied with an address that points to the data memory content, which is then placed on the bus.) (iv) A *control unit* (CU), that receives as input the instruction code from the IR and orchestrates all resources by enabling components to access the internal bus via the control signals *EA* and *EO*. *A/O* conveys to the ALU the operation selection code (conjunction, $A/O = 0$; disjunction, $A/O = 1$). (v) A *program counter* (PC), which supplies the memory with the address of the active instruction. (vi) An *output register* (OR), that allows the processor to transfer the results of a calculation to the output port. The memory is, as usual, implemented off-chip.

Figure 1b depicts the timing diagram of the device, using three clock (CLK) signals. The execution of each instruction occurs in two sequences – a *FETCH* sequence, followed by an execute (*EXE*) sequence. The *FETCH* sequence consists of two phases: in a first phase, the content of the external memory (at the address stored in the PC) is loaded into the IR; the PC is then incremented in a second step. During the *EXE* sequence, which is implemented here in a single phase, the microprocessor decodes and executes the command stored in the IR. This cycle is repeated continuously. Each phase is triggered by a CLK signal (*CLK1*, phase 1; *CLK2*, phase 2; *CLK3*, phase 3). In order to be flexible in terms of clock rate and timing, we generated the CLK signals externally; an on-chip implementation, however, is straightforward. Figure 1c summarizes the instruction set that we have implemented. The instructions are encoded with two bits; some of them are followed by one bit of data. The no-operation (*NOP*) instruction has no effect other than to increase the PC. *LDA* allows the transfer of data from the memory into the AC. *AND* and *OR* perform logical conjunction and disjunction operations, respectively.

It is instructive to consider a simple example. The program fragment

| Address | Mnemonic | OpCode |
|---|---|---|
| 0 | LDA 0 | 010 |
| 1 | AND 1 | 101 |

transfers in a first step, triggered by *CLK1*, the bit sequence 010 from the memory into the IR. *CLK2* then increases the PC and the next instruction becomes available, but is not loaded into the IR yet. Triggered by *CLK3*, the CU then signals the AC (*EA* = 1) to receive the data (0) from the IR via the internal bus. With the next *CLK1* signal, the content of the IR is updated (IR = 101), and the CU enables the ALU to perform a logical conjunction operation (*A/O* = 0) between the data on the bus (1) and that stored in the AC during the previous instruction. Triggered by *CLK3*, the result of this operation (0) is finally written into the OR (*EO* = 1).

We now come to the actual device implementation using a 2D semiconductor. Our microprocessor was fabricated in gate-first technology on a silicon wafer with 280-nm-thick silicon dioxide. The substrate fulfills no other function than acting as a carrier medium and could thus be replaced by glass or any other material, including flexible substrates. We fabricated 18 devices per wafer, with FET channels made from chemical vapor deposition (CVD) grown large-area bilayer $MoS_2$ films. Two Ti/Au metal layers were used to interconnect the transistors and $Al_2O_3$ was used as gate oxide. A detailed description of the device fabrication steps can be found in the Methods section. Subunits, such as e.g. the ALU or the IR, were provided with metal pads for individual testing in a wafer probe station. All subunits were eventually bonded together and the sample was placed back into the probe chamber, where it remained in vacuum for final testing of the complete circuit.

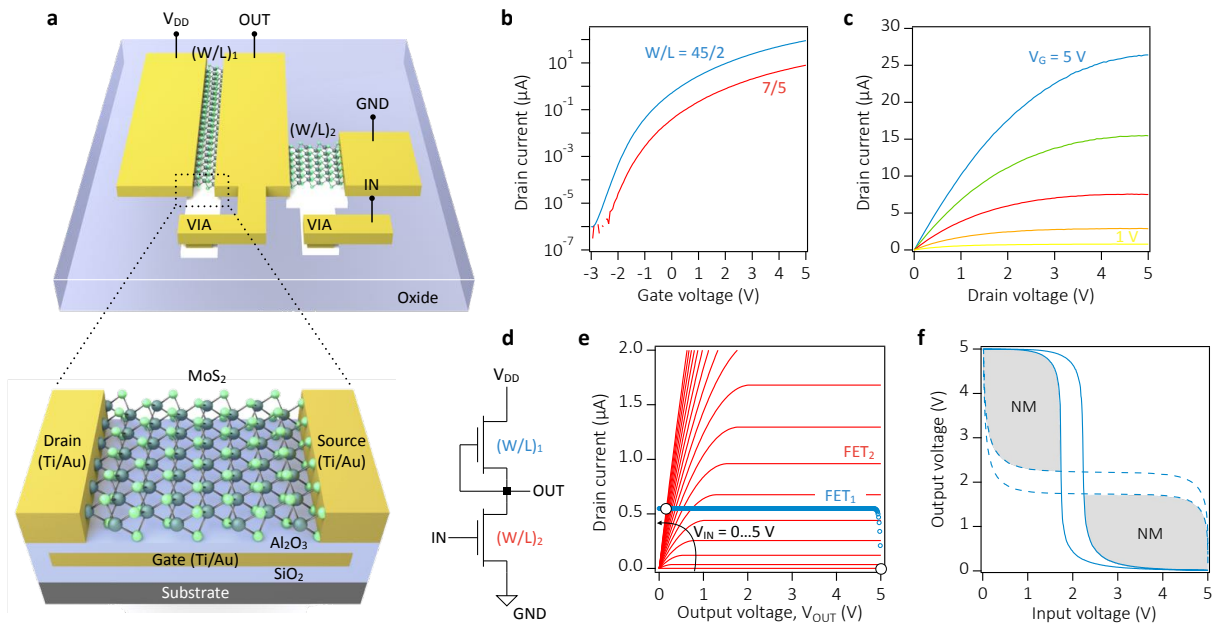

**Figure 2 | Characterization of $MoS_2$ transistors and inverter. (a)** Schematic drawing of an inverter circuit (top) and an individual $MoS_2$ transistor (bottom) in gate-first technology. **(b)** Transfer characteristics of load ($W/L = 45/2$) and pull-down ($W/L = 7/5$) transistors. **(c)** Output characteristic for gate voltages between 1 and 5 V (in 1 V steps). **(d)** NMOS inverter circuit schematic. **(e)** Graphical construction to determine the output voltage $V_{OUT}$ of an inverter for a given input voltage $V_{IN}$. The blue symbols show the load curve and the red lines are the output characteristics of the pull-down transistor (in 0.25 V steps). The intersection point of both curves determines $V_{OUT}$. **(f)** The solid line shows the measured voltage transfer characteristic of an inverter. By mirroring this curve (dashed line) a "butterfly plot" is obtained, from which $NM$ can be extracted by nesting the largest possible square in the grey shaded area.

Figure 2a (bottom) shows a schematic drawing of a so-obtained MoS$_2$ FET. The devices exhibit a field-effect mobility of ~3 cm$^2$/Vs, a threshold voltage $V_T$ of ~0.65 V, an on/off ratio of ~10$^8$, and uniform behavior over a ~50 mm$^2$ area over the wafer (see Supplementary Information). The circuit is based on the NMOS logic family, where both pull-up (load) and pull-down networks were realized using n-type enhancement mode FETs. The implementation of an inverter (see circuit schematic in Figure 2d) using this logic family is shown in Figure 2a (top). A careful design of the $W/L$-ratios, where $W$ and $L$ denote the width and length of the FET channels, is crucial, as it determines the switching threshold and thus the ability to cascade logic stages. $W/L$-ratios of the pull-up and pull-down transistors were made 45/2 (μm/μm) and 7/5, respectively. As demonstrated in the Supplementary Information, the asymmetric design results in improved switching performance as opposed to a symmetric design. Logic NAND gates with $M$ inputs were implemented by connecting $M$ pull-down transistors with $W/L = (M \times 7)/5$ in series. The processor was realized by using a combination of these elements. The minimum feature size of 2 μm was chosen rather large for two reasons. It makes the design immune to sample inhomogeneities (e.g. small holes, cracks, and contaminations in the MoS$_2$ film) and also allows for fast visual inspection of the lithographic structures with an optical microscope.

Figure 2b shows the transfer characteristics of load and pull-down transistors, where the ~14 times higher current through the former demonstrates reliable controllability of the device characteristics by geometrical scaling. The output characteristic, depicted in Figure 2c, shows clear current saturation due to channel pinch-off at the drain. The voltage transfer characteristic of our inverters exhibit excellent performance for a wide supply voltage range between $V_{DD} = 2$ and 7 V, with input and output logic levels being perfectly matched. Figure 2f (solid line) shows the results for $V_{DD} = 5$ V, for which the voltage gain $A_V = -dV_{OUT}/dV_{IN}$ reaches values of $A_V \approx 60$ V/V. Although the voltage transfer curve shows some hysteresis (that mostly stems from trap charges in the gate oxide) the noise margin of the inverter (see shaded area in Figure 2f), $NM \approx 0.59 \times (V_{DD}/2)$, is sufficiently large for integration into multi-stage logic circuits. The NAND gates showed comparable performance. We estimate a static power consumption of $P_s = V_{DD}(I_{D,L} + I_{D,H})/2 \approx 1.4$ μW per logic gate, where $I_{D,L}$ and $I_{D,H}$ denote the currents at $V_{IN} = 0$ and 5 V (see Figure 2e), respectively. The total power consumption of the circuit, consisting of 41 stages, is thus as low as ~60 μW.

A microscope image of the microprocessor is shown in Figure 3a. The device is composed of 115 MoS$_2$ transistors and measures – without bonding pads – 0.6 mm$^2$ in size. Circuit schematics for a D-Latch and the ALU are shown in Figures 3b and 3c, respectively. The complete schematic is presented in the Supplementary Information. A D-Latch is a bi-stable circuit that can be used as 1-bit data storage element, triggered by a CLK signal. It forms the basic building block of all our data registers (IR, AC, and OR) and the PC. The ALU is a combinational logic circuit, entirely based on NANDs, that performs bitwise logic operations on 1-bit data. The additional input *A/O* signals the ALU which operation to perform.

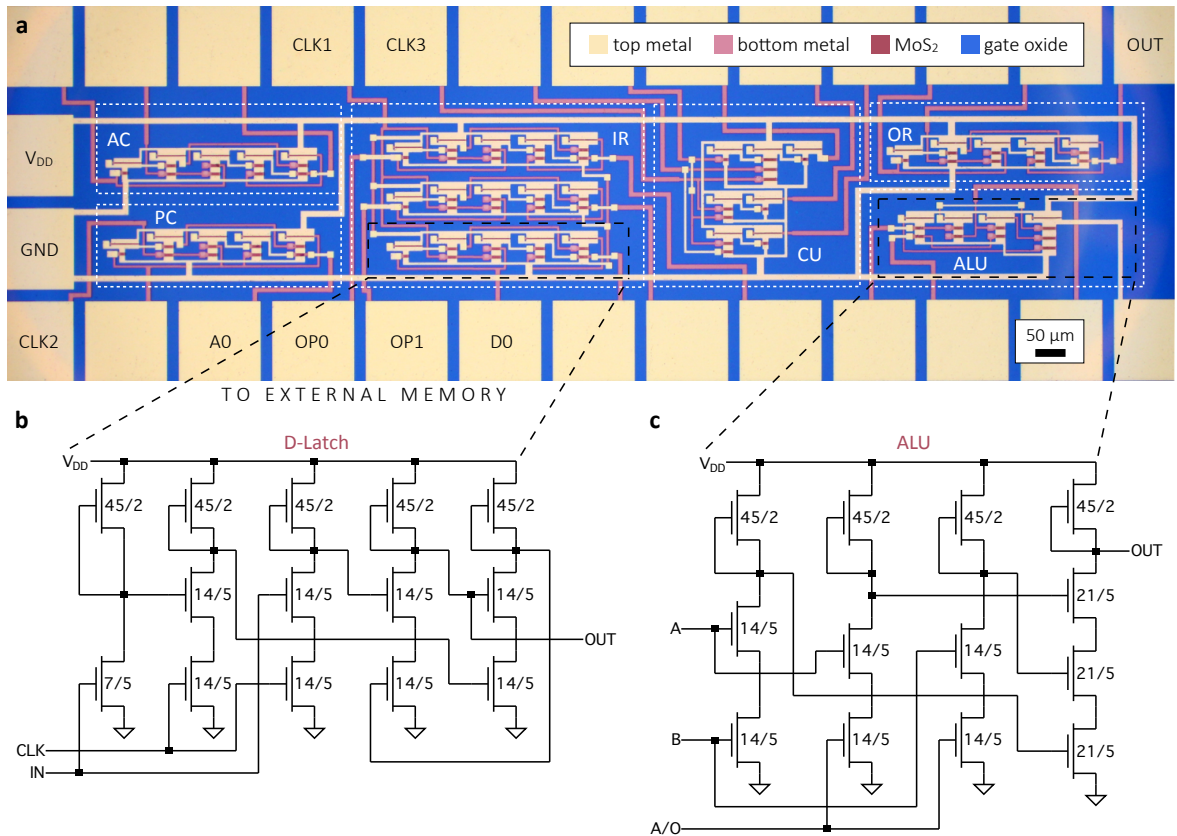

**Figure 3 | Device implementation using a 2D semiconductor. (a)** Microscope image of the microprocessor. The two metal layers appear in different color and are connected with via-holes. All subunits were provided with metal pads for individual testing. Labeled pads were used to connect the device to the periphery (memory, CLK signal generation, power supply, output), the others were wire bonded together to realize the internal connections. Scale bar, 50 μm. Circuit schematics of **(b)** D-Latch and **(c)** ALU, with $W/L$-ratio in units of μm/μm for each transistor. *IN*, input; *OUT*, output. The complete microprocessor schematic is presented in the Supplementary Information.

We first verified the functionality of the microprocessor by running the example program from above and measuring waveforms at different locations on the chip (see Methods for measurement details). As shown in Figure 4a, the device is indeed able to deliver the correct result, with excellent signal integrity and with rail-to-rail performance, proving the ability to cascade logic stages based on 2D semiconductors. To further demonstrate the operability of the device, we present in Figure 4b the results from a series of logical disjunction operations. The match of measured and expected outputs shows again correct operation. As shown in the Supplementary Information, the device proved to be functional at CLK frequencies of 50 Hz. This is by no means a limitation of the TMD material itself, but is caused by the limitations of our measurement setup. Ultimately, the speed is limited by the current-driving capability of the pull-up transistor, which is operated in the sub-threshold regime ($V_{GS} = 0 < V_T$) and acts as current source with $I_D \approx 0.55$ µA. For a typical (external) capacitive load of $C_L \approx 1$–$10$ pF, we estimate a maximum operation frequency of $f_{MAX} \approx I_D/(2\pi V_{DD} C_L) \approx 2$–$20$ kHz. To increase $f_{MAX}$, $I_D$ could be increased by employing depletion mode load FETs [20], by controlled chemical doping, or just by improving the carrier mobility of the 2D semiconductor.

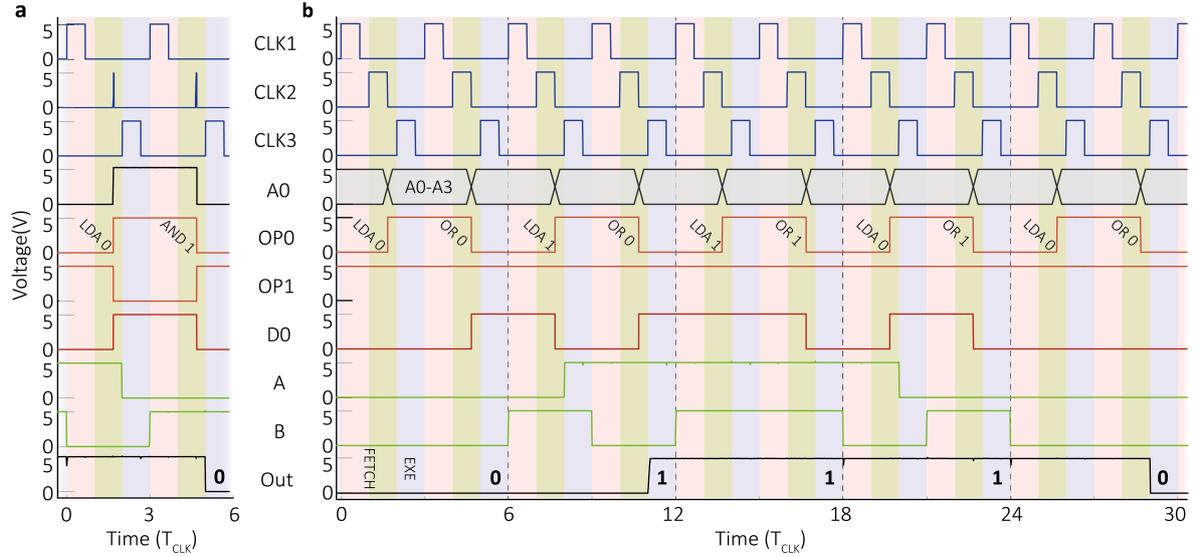

**Figure 4 | Device operation. (a)** Waveforms measured on the chip when running the sample program, explained above. The *CLK1*, *CLK2*, and *CKL3* signals were generated externally. Each instruction requires three CLK cycles, where $1/T_{CLK}$ is the CLK frequency. *CLK2* pulses were made sufficiently short to trigger the level-sensitive input of the PC; this could be avoided by using a more complex master-slave design. *A0* is the address supplied by the PC. *OP0*, *OP1*, and *D0* denote the signals from the memory, where the former two are the instruction and the latter is the data. *A* and *B* are the input signals to the ALU, and *OUT* is the output of the device. $T_{CLK} = 500$ ms. **(b)** Results for a series of other calculations. In order to run the longer program, a 4-bit PC was implemented externally. The meaning of the curves is the same as in **a**, apart from *A0* which schematically depicts here the 4-bit address signal (*A0–A3*). Again, the device is able to deliver the expected logic values, shown as numbers at the bottom.

In summary, we have reported a first step towards the development of microprocessors based on 2D semiconductors. The major challenge that we faced during device fabrication is yield. Although the yield for subunits was high (e.g. ~80 % of ALUs were fully functional), the sheer complexity of the full system, together with the non-fault tolerant design, resulted in an overall yield of only a few percent. Imperfections of the MoS$_2$ film, mainly caused by the transfer from the growth to the target substrate, were identified as main source for device failure. However, as no metal catalyst is required for the synthesis of TMD films [27–30], direct growth on the target substrate is a promising route to improve yield. Besides that, we do not see any roadblocks that could prevent the scaling of our 1-bit design to multi-bit data. Our work demonstrates that integrated circuits consisting of 2D materials are a promising emerging technology.

**METHODS**

Device fabrication started with patterning of the bottom metal (gate) layer by electron beam lithography (EBL) and evaporation of Ti/Au (5/25 nm). A 22-nm-thick $Al_2O_3$ gate oxide was then deposited using atomic layer deposition (ALD), followed by a second lithography step and wet chemical etching in potassium hydroxide (KOH) to define the via-holes that connect the bottom and top metal layers where necessary. Following the procedure described in Ref. [29], a large-area $MoS_2$ film was grown by chemical vapor deposition on sapphire and then transferred onto the target wafer. The film is continuous over an area of ~50 $mm^2$ with bilayer thickness and small multi-layer $MoS_2$ islands and contaminations. The $MoS_2$ film was characterized by atomic force microscopy and Raman spectroscopy (see Supplementary Information). In a third EBL step, rectangular $MoS_2$ channels were patterned and subsequently etched using Ar/SF6 plasma. Before lift-off, mild treatment of the sample in oxygen plasma was performed to remove the crust from the surface of the polymer mask. The top metal (drain/source contact) layer was then formed by another EBL process and subsequent Ti/Au (5/35 nm) deposition. The sample was finally annealed in vacuum at 400 K for several hours to remove adsorbants from the surface and reduce device hysteresis. For testing, we generated the CLK signals externally, using a digital I/O card in a computer. The same card was used for emulating the external memory. The device was supplied with $V_{DD} = 5$ V, and waveforms were recorded with a Semiconductor Parameter Analyzer, connected to the probe tips of a wafer probe station.

**Acknowledgments:** We thank Simone Schuler, Andreas Pospischil, Marco Furchi, Andreas Kleinl, Fabian Doná, Werner Schrenk, Markus Schinnerl, Peter Kröll, and Benedikt Gottsbachner for technical assistance, Alois Lugstein and Emmerich Bertagnolli for providing access to CVD and ALD systems, and Dumitru Dumcenco for helpful discussions. We acknowledge financial support by the Austrian Science Fund FWF (START Y 539-N16) and the European Union (grant agreement No. 696656 Graphene Flagship).

**Author contributions:** T.M. conceived the experiment. S.W. and T.M. designed the circuit. D.K.P. grew the $MoS_2$ films. S.W. and D.K.P. fabricated the devices and carried out the measurements. O.B. contributed to the sample fabrication. T.M. prepared the manuscript. All authors discussed the results and commented on the manuscript.

**Competing financial interests:** The authors declare no competing financial interests.

# REFERENCES

1. Cooke, M. High-mobility channels and moving beyond silicon. *Semiconductor Today* **10**, 88–91 (2015).
2. Myny, K. *et al.* An 8-Bit, 40-Instructions-Per-Second Organic Microprocessor on Plastic Foil. *IEEE J. Solid-State Circuits* **47**, 284–291 (2012).
3. Shulaker, M.M. *et al.* Carbon nanotube computer. *Nature* **501**, 526–530 (2013).
4. Mak, K.F., Lee, C., Hone, J., Shan, J. & Heinz, T.F. Atomically Thin $MoS_2$: A New Direct-Gap Semiconductor. *Phys. Rev. Lett.* **105**, 136805 (2010).
5. Radisavljevic, B., Radenovic, A., Brivio, J., Giacometti, V. & Kis, A. Single-layer $MoS_2$ transistors. *Nature Nanotech.* **6**, 147–150 (2011).
6. Li, L. *et al.* Black phosphorus field-effect transistors. *Nature Nanotech.* **9**, 372–377 (2014).
7. Tao, L. *et al.* Silicene field-effect transistors operating at room temperature. *Nature Nanotech.* **10**, 227–231 (2015).
8. Yan, R.H., Ourmazd, A. & Lee, K.F. Scaling the Si MOSFET: from Bulk to SOI to Bulk. *IEEE Trans. Electron Dev.* **39**, 1704–1710 (1992).
9. Fiori, G. *et al.* Electronics based on two-dimensional materials. *Nature Nanotech.* **9**, 768–779 (2014).
10. Liu, H., Neal, A.T. & Ye, P.D. Channel Length Scaling of $MoS_2$ MOSFETs. *ACS Nano* **6**, 8563–8569 (2012).
11. Ilatikhameneh, H. *et al.* Tunnel Field-Effect Transistors in 2-D Transition Metal Dichalcogenide Materials. *IEEE J. Explor. Solid-State Computat. Devices Circuits* **1**, 12–18 (2015).
12. Roy, T. *et al.* Dual-Gated $MoS_2$/$WSe_2$ van der Waals Tunnel Diodes and Transistors. *ACS Nano* **9**, 2071–2079 (2015).
13. Pospischil, A. & Mueller, T. Optoelectronic Devices Based on Atomically Thin Transition Metal Dichalcogenides. *Appl. Sci.* **6**, 78 (2016).
14. Pu, J. *et al.* Highly Flexible $MoS_2$ Thin-Film Transistors with Ion Gel Dielectrics. *Nano Lett.* **12**, 4013–4017 (2012).
15. Lee, G.-H. *et al.* Flexible and Transparent $MoS_2$ Field-Effect Transistors on Hexagonal Boron Nitride-Graphene Heterostructures. *ACS Nano* **7**, 7931–7936 (2013).
16. Cheng, R. *et al.* Few-layer molybdenum disulfide transistors and circuits for high-speed flexible electronics. *Nature Comm.* **5**, 5143 (2014).
17. Podzorov, V., Gershenson, M.E., Kloc, Ch., Zeis, R. & Bucher, E. High-mobility field-effect transistors based on transition metal dichalcogenides. *Appl. Phys. Lett.* **84**, 3301–3303 (2004).
18. Ayari, A., Cobas, E., Ogundadegbe, O. & Fuhrer, M.S. Realization and electrical characterization of ultrathin crystals of layered transition- metal dichalcogenides. *J. Appl. Phys.* **101**, 014507 (2007).
19. Radisavljevic, B., Whitwick, M.B. & and Kis, A. Integrated Circuits and Logic Operations Based on Single-Layer $MoS_2$. *Nano Lett.* **5**, 9934–9938 (2011).
20. Wang, H. *et al.* Integrated Circuits Based on Bilayer $MoS_2$ Transistors. *Nano Lett.* **12**, 4674–4680 (2012).
21. Yu, L. *et al.* Design, Modeling and Fabrication of CVD Grown $MoS_2$ Circuits with E-Mode FETs for Large-Area Electronics. *Nano Lett.*, DOI: 10.1021/acs.nanolett.6b02739 (2016).
22. Tosun, M. *et al.* High-Gain Inverters Based on $WSe_2$ Complementary Field-Effect Transistors. *ACS Nano* **8**, 4948–4953 (2014).

23. Yu, L. *et al.* High-Performance WSe$_2$ Complementary Metal Oxide Semiconductor Technology and Integrated Circuits. *Nano Lett.* **15**, 4928–4934 (2015).
24. Kim, S. *et al.* High-mobility and low-power thin-film transistors based on multilayer MoS$_2$ crystals. *Nature Comm.* **3**, 1011 (2012).
25. Das, S., Chen, H.-Y., Penumatcha, A.V. & Appenzeller, J. High Performance Multilayer MoS$_2$ Transistors with Scandium Contacts. *Nano Lett.* **13**, 100–105 (2013).
26. Schwierz, F., Pezoldt, J. & Granzner, R. Two-dimensional materials and their prospects in transistor electronics. *Nanoscale* **7**, 8261–8283 (2015).
27. Liu, K.-K. *et al.* Growth of Large-Area and Highly Crystalline MoS$_2$ Thin Layers on Insulating Substrates. *Nano Lett.* **12**, 1538–1544 (2012).
28. Zhan, Y., Liu, Z., Najmaei, S., Ajayan, P.M. & Lou, J. Large-area vapor-phase growth and characterization of MoS$_2$ atomic layers on a SiO$_2$ substrate. *Small* **8**, 966–971 (2012).
29. Dumcenco, D. *et al.* Large-Area Epitaxial Monolayer MoS$_2$. *ACS Nano* **9**, 4611–4620 (2015).
30. Kang, K. *et al.* High-mobility three-atom-thick semiconducting films with wafer-scale homogeneity. *Nature* **520**, 656–660 (2015).

*Supplementary Information for*

# A microprocessor based on a two-dimensional semiconductor

**Stefan Wachter**[1,*]**, Dmitry K. Polyushkin**[1,*]**, Ole Bethge**[2]**, and Thomas Mueller**[1,‡]

[1] *Vienna University of Technology, Institute of Photonics, Gußhausstraße 27-29, 1040 Vienna, Austria*
[2] *Vienna University of Technology, Institute of Solid State Electronics, Floragasse 7, 1040 Vienna, Austria*

[*] These authors contributed equally to this work.
[‡] Corresponding author: thomas.mueller@tuwien.ac.at

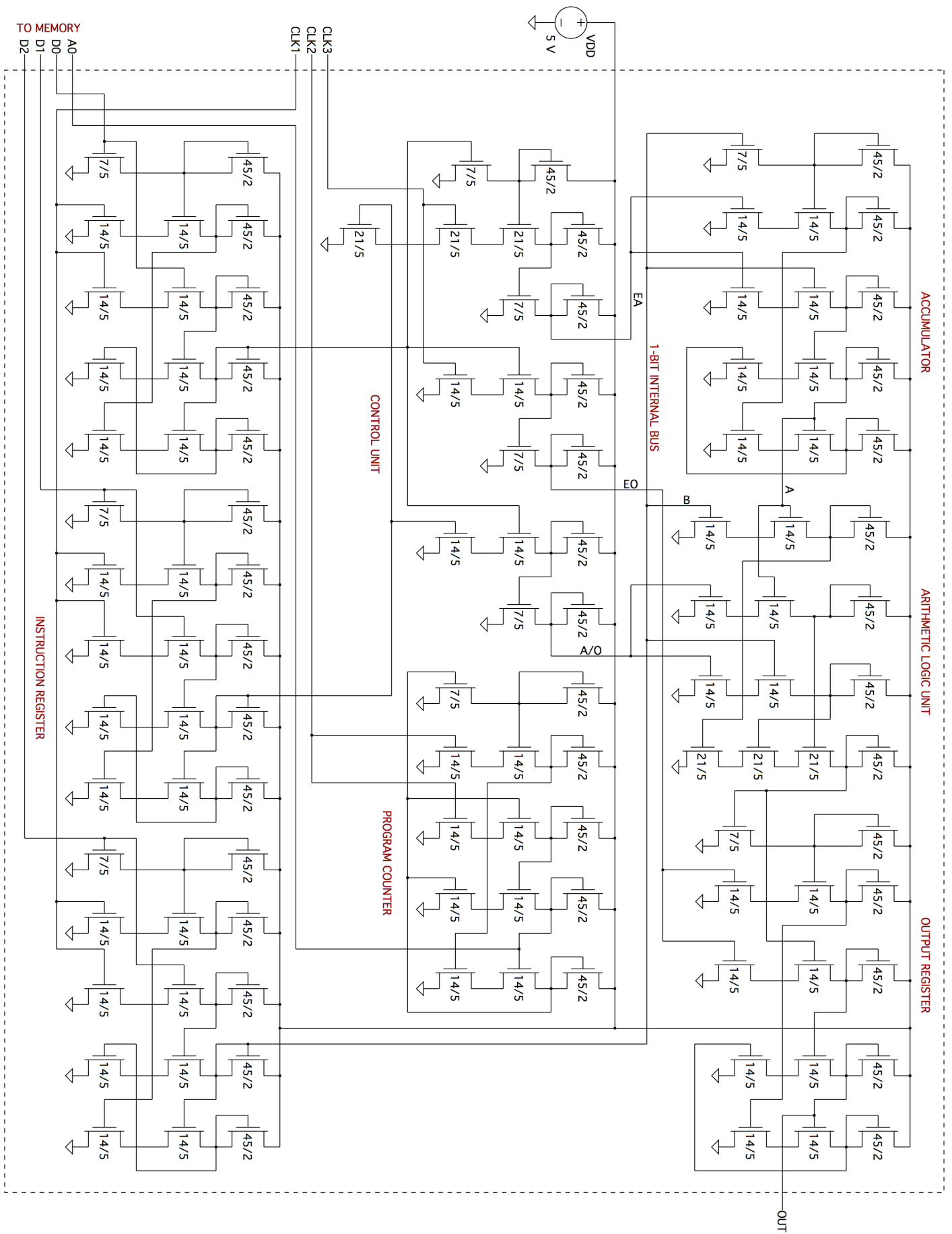

*Supplementary Figure 1.* Circuit schematic of the microprocessor.

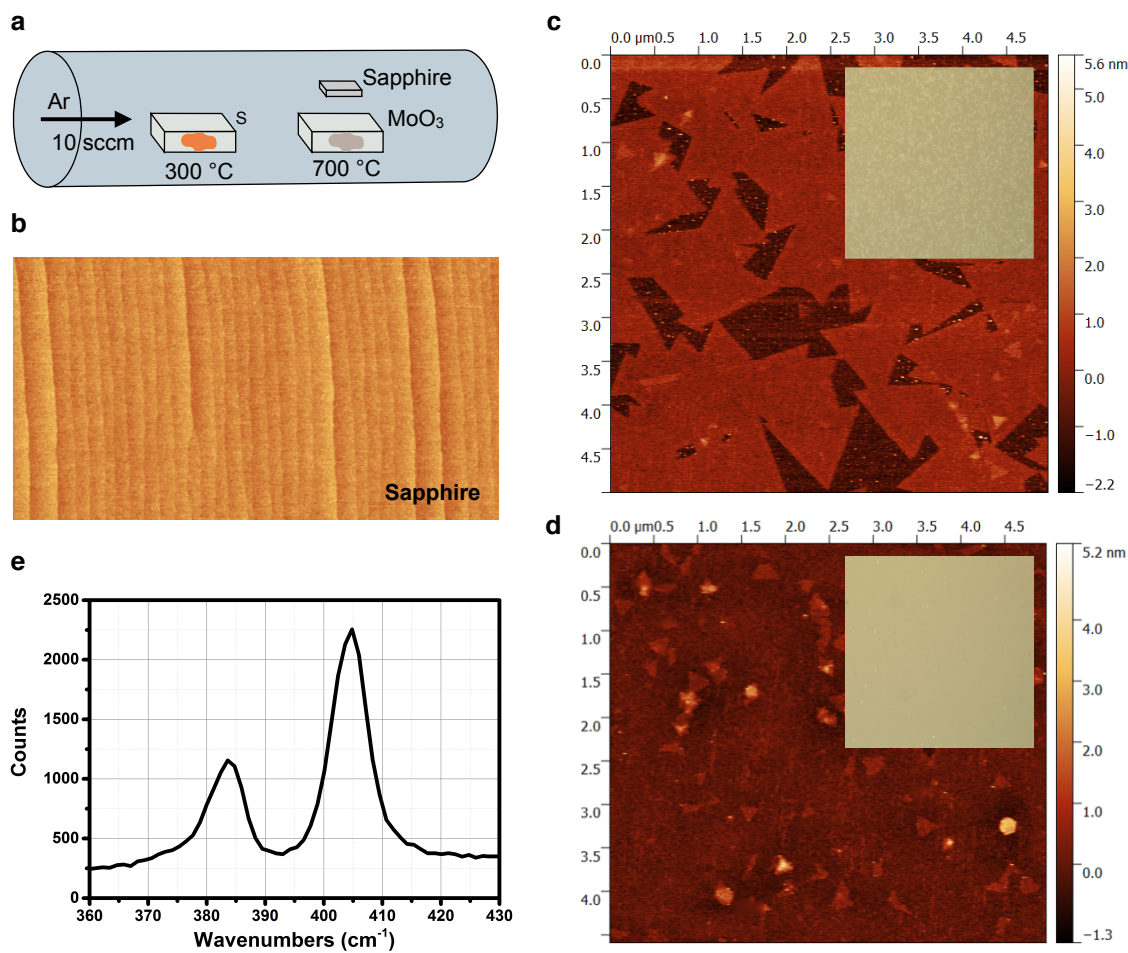

*Supplementary Figure 2.* MoS₂ film growth. (a) Schematic illustration of the growth setup. (b) AFM image of sapphire growth substrate. (c) The AFM image, taken close to the edges of the sample, shows triangular MoS₂ monolayer growth (inset: microscope image of the discontinuous MoS₂ film). (d) Towards the center of the sample, the triangles coalesce and form a continuous, polycrystalline film. (e) Raman measurements reveal that the MoS₂ film in the center is a bilayer ($E_{2g} - A_{1g}$ spacing: 21.2 cm⁻¹).

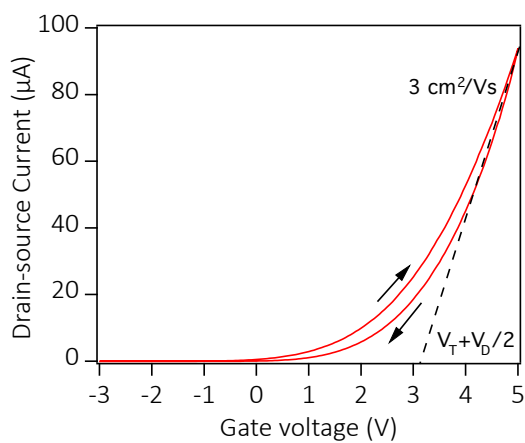

*Supplementary Figure 3.* Field-effect mobility ($\mu \approx 3$ cm²/Vs), hysteresis, and threshold voltage ($V_T \approx 0.65$ V) of MoS₂ FET. $V_D = 5$ V.

*Supplementary Note 1: Inverter input-output characteristic.* For simple analytic modeling, we performed calculations based on long-channel FET theory. The pull-down FET is described by $I_{D2} = K_2[(V_{IN} - V_T)V_{OUT} - V_{OUT}^2/2]$ in the triode regime and $I_{D2} = K_2(V_{IN} - V_T)^2/2$ in the saturation regime (red curves in Figure 2e). The load FET is operated in the sub-threshold regime ($V_{G1} = 0 < V_T$), and thus acts as a current source over a large drain voltage range, $I_{D1} = K_1[1 - \exp(-\beta V_{D1})]$ with $\beta$ being the reciprocal of the thermal potential (Sze, *Physics of Semiconductor Devices*). From the circuit schematic Figure 2d, it is apparent that $V_{D1} = V_{DD} - V_{OUT}$, and thus $I_{D1} = K_1[1 - \exp(\beta V_{OUT} - \beta V_{DD})]$ (blue symbols in Figure 2e). By equating both currents, $I_{D1} = I_{D2}$, we obtain a relation between $V_{OUT}$ and $V_{IN}$. The parameters $K_1$ and $K_2$ are taken from the experiment (Figure 2b). Supplementary Figure 4a shows the results for the asymmetric transistor design presented in the main manuscript (Figure 2a). If both transistors are implemented with same $W/L$-ratio, the switching threshold drops below 1 V (Supplementary Figure 4b), resulting in low noise margin (especially in the presence of additional hysteresis).

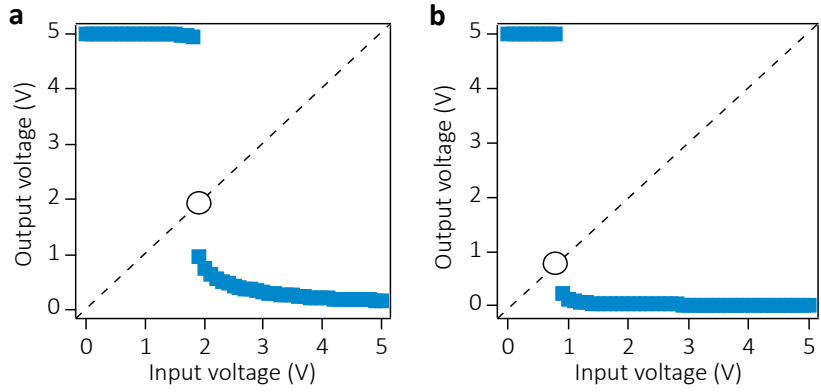

*Supplementary Figure 4.* Transfer curves for (a) asymmetric and (b) symmetric design.

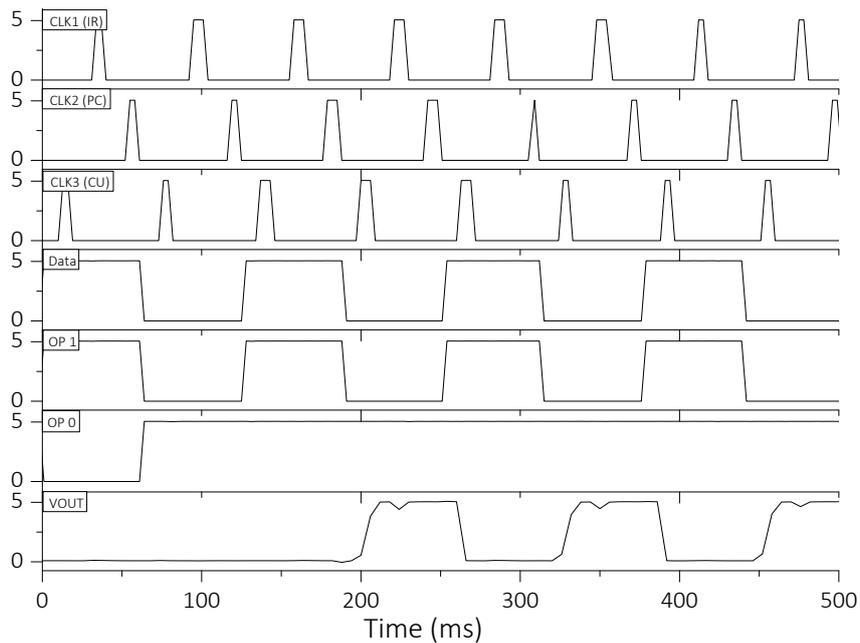

*Supplementary Figure 5.* Device operation at $1/T_{CLK} = 50$ Hz.